# *i*MOD LEACH: improved MODified LEACH Protocol for Wireless Sensor Networks

S. Ahmed, M. M. Sandhu, N. Amjad, A. Haider, M. Akbar, A. Ahmad, Z. A. Khan*, U. Qasim[#], N. Javaid

COMSATS Institute of Information Technology, Islamabad, Pakistan.
*Internetworking Program, Faculty of Engineering, Dalhousie University, Halifax, Canada.
[#]University of Alberta, Alberta, Canada.

**ABSTRACT**

Increased use of Wireless sensor Networks (WSNs) in variety of applications has enabled the designers to create autonomous sensors, which can be deployed randomly, without human supervision, for the purpose of sensing and communicating valuable data. Many energy-efficient routing protocols are designed for WSNs based on clustering structure. In this paper, we have proposed *i*MODLEACH protocol which is an extension to the MODLEACH protocol. Simulation results indicate that *i*MODLEACH outperforms MODLEACH in terms of network life-time and packets transferred to base station. The mathematical analysis helps to select such values of these parameters which can suit a particular wireless sensor network application.

**INDEX TERMS**— Wireless Sensor Networks, Routing Protocols, Clustering, Base Station, Cluster Head.

## 1. INTRODUCTION

Wireless Sensor Networks offer unique benefits and versatility in terms of low-power and low-cost rapid deployment for applications which do not require human supervision. Nodes in WSNs are usually battery operated sensing devices with limited energy resources. Thus energy efficiency is one of the most important issues and designing power-efficient protocols is critical for prolonging the lifetime. WSNs have been considered for certain applications with limited power, reliable data transfer, short range communication, and reasonably low cost such sensing applications [1].

Many energy-efficient routing protocols are designed based on clustering structure. The clustering technique can also used to perform data aggregation, which combines data from source nodes into set of meaningful information [2]. Many routing protocols have originated since the development of this field in which LEACH [3], DEEC [2], TEEN, SEP and PEGASIS are some of them. LEACH, of them, proved to be more promising and became a bench-mark in the designing of other protocols like A-sLEACH [4], Enhanced LEACH [5], LEACH-CC [6], Ad-LEACH [7] and MODLEACH [8] are some of them.

In this paper, we have considered MODLEACH protocol as the reference and considered such parameters like *p* (probability of choosing a CH), *s* (software threshold) and *h* (hard threshold) which have been used in this protocol to further enhance the performance of the MODLEACH protocol. As our protocol out-performs the functionality of the protocol, hence the name is given as *i*MODLEACH (Improved Modified LEACH). We varied the values of the *p*, *s* and *h* parameters and analyzed their effect on the performance of the network mathematically and then verified those with the help of simulations. The results clearly indicate that our protocol *i*MODLEACH outperforms MODLEACH in terms of network life-time and packets transferred to base station. The analysis helps to select such values of these parameters which can suit a particular wireless sensor network application.

## 2. Related Work

Limited energy resources of sensor nodes create challenging issues on the improvement of routing protocols for WSN. Introducing clustering into network's topology reduces number of transmissions in the network; hence providing energy efficiency as CHs aggregate data from their respective member nodes as well as reducing replica of transmission and enhancing the network lifetime.

Heinzelman, et.al [3] introduced a clustering algorithm for sensor networks, called Low Energy Adaptive Clustering Hierarchy (LEACH). LEACH forms clusters by using a distributed algorithm, where nodes make autonomous decisions without any centralized control. LEACH arranges the nodes in the network into clusters and chooses one of them as CH. The operation of LEACH is divided into rounds. Each round begins with a setup phase when the clusters are organized, followed by a steady-state phase when data is transferred from nodes to the CH and then to the BS.

LEACH forms the bases for many protocols and many flavors have come up introducing many variations in the basic algorithm each having its own beauty and improvements. In [4], Junayed Islam et al. presented a A-sLEACH (An Advanced Solar Aware Leach Protocol for Energy Efficient Routing in WSNs), a clustering based protocol which introduced the idea of sensor radio model for randomization of local CHs. It enhanced data aggregation by FIFO priority scheme and collision minimized non-persistent Carrier Sense Multiple Access (CSMA). In [5], the authors proposed a clustering routing protocol by the name of Enhanced LEACH, which extended LEACH protocol by balancing energy consumption in the network. Their simulation results show that Enhanced LEACH outperforms LEACH in terms of network lifetime and power consumption minimization. Again in [6], a low energy-consumption chain-based routing protocol LEACH-CC was proposed. The new protocol is characterized by each node will send information about its current location and energy level to BS. BS runs the simulated annealing algorithm to determine the clusters for that round. A chain routing is established between clusters to reduce the amount of nodes which communicate with the BS. The experimental results show that LEACH-CC outperforms in terms of network lifetime and energy efficiency. In [7], authors have introduced an Ad-LEACH static clustering based heterogeneous routing protocol with a cluster head selection technique adopted from DEEC [2]. It enhances both LEACH and DEEC protocols both in terms of energy efficiency and throughput. In [8], D.Mehmood et.al has given a MODLEACH protocol by introducing efficient cluster head replacement scheme and dual transmitting power levels. Authors in [9] proposed Density Controlled Divide-and-Rule LEACH. This protocol selects optimal number of CHs on the bases of nodes' density such that uniform distribution of load on CHs is maintained throughout the network operation. Similarly, [10] solves the problem of non uniform load distribution on CHs by removing away CHs. In [11], authors proposed Hybrid Energy Efficient Reactive protocol for WSNs. Basically, this protocol aims to maximize network lifetime by minimized transmissions in a heterogeneous environment. Authors in [12] introduce a routing mechanism which carries the network operation in such a way that the network lifetime is enhaced due to their energy hole removal mechanism. In [13], authors addressed the issues related to throughput maximization and delay minimization, and suggested a linear programming based solution. On the other hand, authors in [14] conducted a comprehensive study on the evaluation of LEACH based clustering protocols, whereas, evaluation of routing protocols on the bases of scalability and traffic constraints is performed in [15]. Moreover, in [16] authors analyzed multi level Hierarchal routing protocols under the umbrella of energy efficiency through proper route selection.

## 3. Problem Definition

Heinzelman, et al. [3] Low Energy Adaptive Clustering Hierarchy (LEACH) is a cluster-based protocol, in which the role of a member node in the cluster is to sense the surrounding environment and transmit the sensed data to a cluster head. The cluster head then aggregates and transfers the sensed data to the BS in order to reduce the amount of information that must be transmitted to the BS, which drains out more energy of the cluster head as compared to the other nodes. The protocol randomly selects a few sensor nodes as cluster-heads and rotates this role to evenly distribute the energy load among the sensors in the network.

LEACH uses a TDMA/code-division multiple access (CDMA) MAC to reduce inter-cluster and intra-cluster collisions. After a given epoch, randomized rotation of the role of cluster-head is conducted so that uniform energy dissipation in the sensor network is obtained. A parameter of p is introduced which reflects the probability of choosing a cluster head among the member nodes of a cluster.

Several flavors of LEACH have cropped up since its origin and each one of them has shown certain improvements in different features and parameters of LEACH. MODLEACH [8] is one of those protocols but this protocol, like other protocols, has not considered the role of p on the performance of the protocol, mainly the stability of the network, packets transferred to CH and BS etc. Also MODLEACH introduced a soft threshold and hard threshold parameters and assigned them a fixed value. Rather varying the values of p as well as h parameters certainly shows considerable improvements in the performance of the network. In this research, certain mathematical analysis was done to find the role of these parameters and to come up with certain nominal values which must be selected while considering the application domain of the network. Experiments were done in MATLAB to find the accuracy of the mathematical analysis done as described in the sections below.

## 4. Proposed *i*MODLEACH Protocol

In this section we will discuss the *i*MODLEACH protocol in detail. This protocol is primarily based on MODLEACH [8] protocol which is itself an inspiration of the LEACH [3] protocol.

The type of network is heterogeneous. There are n numbers of nodes, randomly distributed across region. The main region is further divided into sub regions which are normally referred to as, clusters. Each cluster contains number of nodes of which one of them is acting as the CH. Each CH receives data from all of its client nodes and performs some necessary iteration for compression. All CH's forward the compressed data to Base Station. All nodes are considered nomadic or stationary within their respective cluster and hence, there is no abrupt change in network topology.

In this research, three very important parameters *p* (probability of choosing a CH), *s* (soft threshold) and *h* (hard threshold) are considered and their impact on net performance of the network are studied and analyzed; both analytically and with simulation. MODLEACH utilized these parameters by selecting their values fixed at *p*=0.1, *s*=2 and *h*=100. We made certain variations in all these parameters and studied their behavior on the performance of the network and other parameters.

First of all varying the value of *p* from 0.1 to 0.9, different experiments were performed in MATLAB and the following readings were noted:

TABLE I
Variation of *p* and its effect on Stability and Network Life-time

| S.No | *p* | Maximum rounds traversed | Packets sent to BS | Packets sent to CH | First dead node at round |
|---|---|---|---|---|---|
| 1 | 0.1 | 1095 | 6985 | 55296 | 160 |
| 2 | 0.3 | 1362 | 16941 | 38880 | 62 |
| 3 | 0.4 | 1528 | 20061 | 39213 | 52 |
| 4 | 0.5 | 1537 | 24433 | 24256 | 36 |
| 5 | 0.6 | 1641 | 22457 | 22116 | 35 |
| 6 | 0.8 | 1930 | 32715 | 8037 | 23 |
| 7 | 0.9 | 2121 | 32956 | 3716 | 21 |

With the help of these results, we can infer that there is a trade-off between the maximum round traversed and the first dead node of the network or in other words the stability of the system, on selection of a particular value of $p$; and hence we come to the following conclusions:

$$Probability\ of\ choosing\ a\ CH\ \propto\ maximum\ round\ of\ a\ network \quad (1)$$

$$Probability\ of\ choosing\ a\ CH \propto \frac{1}{first\ dead\ node\ of\ a\ network} \quad (2)$$

$$Probability\ of\ choosing\ a\ CH \propto packets\ sent\ to\ BS \quad (3)$$

$$Probability\ of\ choosing\ a\ CH \propto \frac{1}{packets\ sent\ to\ cluster\ head} \quad (4)$$

Combining equations (1) and (2) we get;

$$Probability\ of\ choosing\ CH\ =\ k_1 \times \frac{maximum\ rounds\ of\ a\ network}{first\ dead\ node\ of\ a\ network} \quad (5)$$

Combining equations (3) and (4) we get;

$$Probability\ of\ choosing\ CH\ =\ k_2 \times \frac{packets\ sent\ to\ base\ station}{packets\ sent\ to\ cluster\ head} \quad (6)$$

where $k_1$ and $k_2$ are the constants of proportionalities having the values of $0 < k_1, k_2 \leq 1$.

In the second set of experiments, we varied the values of soft threshold from $s=1$ to $s=7$, and observed that no considerable change in the performance of the network was observed.

In the final set of experiments, we varied the values of hard threshold $h$ by keeping the $p$ constant and then making $p$ variable, the experiments were repeated again. The following set of data was generated on this basis:

TABLE II
Variation of $p$ and $h$ and its effect on Stability and Network Life-time

| S.No | $p$ | $h$ | Maximum rounds in a network | First dead node of the network | Ratio x= First dead node/ maximum rounds |
|---|---|---|---|---|---|
| 1 | 0.1 | 100 | 1095 | 160 | 0.146 |
| 2 | 0.1 | 200 | 1248 | 148 | 0.123 |
| 3 | 0.1 | 300 | 1200 | 148 | 0.118 |
| 4 | 0.1 | 400 | 1261 | 131 | 0.104 |
| 5 | 0.1 | 500 | 1106 | 162 | 0.146 |
| 6 | 0.1 | 600 | 1216 | 159 | 0.131 |
| 7 | 0.1 | 700 | 1102 | 143 | 0.129 |
| 8 | 0.1 | 800 | 1207 | 133 | 0.110 |
| 9 | 0.2 | 100 | 1313 | 103 | 0.0784 |
| 10 | 0.2 | 200 | 1503 | 88 | 0.0585 |
| 11 | 0.2 | 300 | 1259 | 98 | 0.0778 |
| 12 | 0.2 | 400 | 1371 | 83 | 0.0605 |
| 13 | 0.2 | 500 | 1231 | 86 | 0.0698 |
| 14 | 0.2 | 600 | 1262 | 88 | 0.0697 |
| 15 | 0.2 | 700 | 1244 | 82 | 0.0565 |
| 16 | 0.2 | 800 | 1255 | 71 | 0.0431 |

The table reveals that for a fix value of $p=0.1$, the ratio of first dead node round and that of the maximum rounds decreases constantly by varying the value of h from 100 to 400, it again rises to the same value at $h=500$ as that of $h=100$ and then decreases again till $h=800$ and the same behavior is repeated for higher values of $h$ and fixing the value of p at some other standard. Hence the selection of h at 100 or 500 gives us the same performance but at values of 200, 300 and 400, we find a compromise again between the stability of the network and the life-time of the network which is again focused on the application in which the network is desired.

Hence, taking into consideration all these facts and figures, we can properly adjust the values of parameters *p*, *s* and *h* and certainly the performance of the MODLEACH is considerably enhanced for different environments of applications.

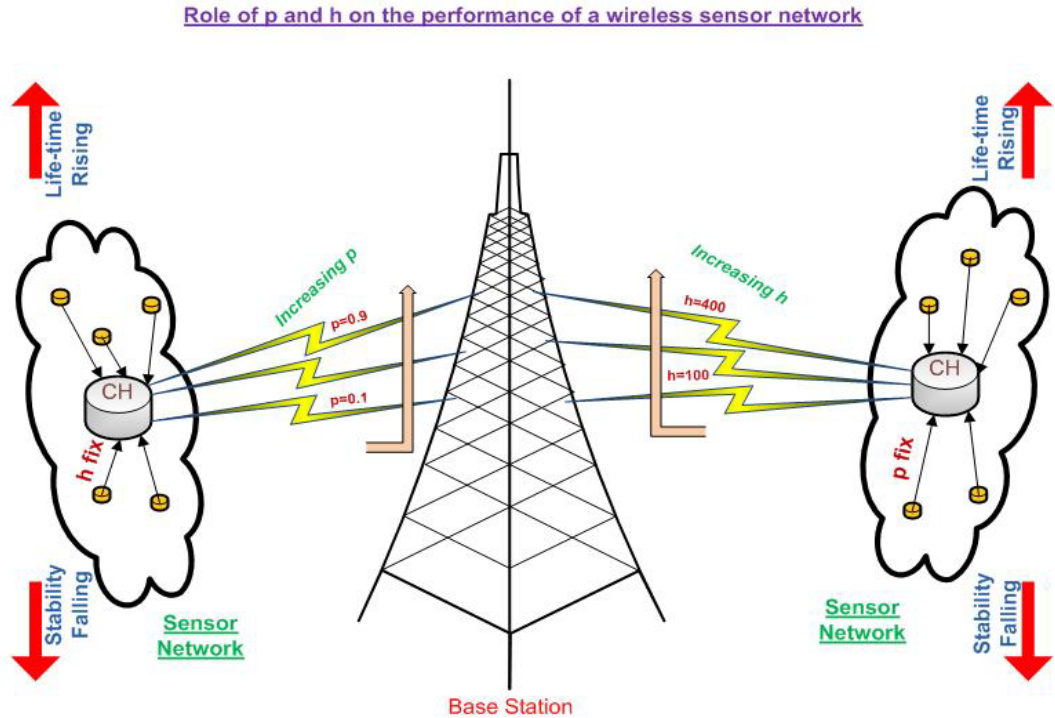

Fig. 1 Basic Theme for the *i*MODLEACH Protocol

### 5. Simulation Results

The mathematical analysis was further verified using simulations on MATLAB in which the following parameters were set and a variety of graphs were plotted which proved that our protocol iMODLEACH outperforms MODLEACH protocol in certain respects.

TABLE III
Network Parameters with Specified Values

| S.No | Network Parameters | Values |
|---|---|---|
| 1 | Network Size | 400 x 400 m$^2$ |
| 2 | Number of nodes | 100 |
| 3 | Sensor nodes initial energy | 0.5 J |
| 4 | Packet Size | 4000 bits |
| 5 | Energy consumption in idle state | 50 nJ/bit |
| 6 | Data aggregation energy consumption | 5 nJ/bit/report |
| 7 | Amplification energy (cluster to BS), d$\leq$ d$_0$, $E_{mp}$ | 0.0013 pJ/bit/m$^2$ |
| 8 | Amplification energy (cluster to BS), d$\geq$ d$_0$, $E_{fs}$ | 10 pJ/bit/m$^2$ |
| 9 | Amplification energy (intra cluster communication), d$\leq$ d$_1$ | $E_{mp}/10 = E_{mp1}$ |
| 10 | Amplification energy (intra cluster communication), d$\geq$ d$_1$ | $E_{fs}/10 = E_{fs1}$ |

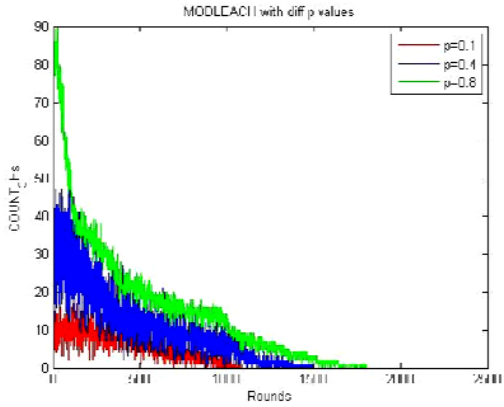
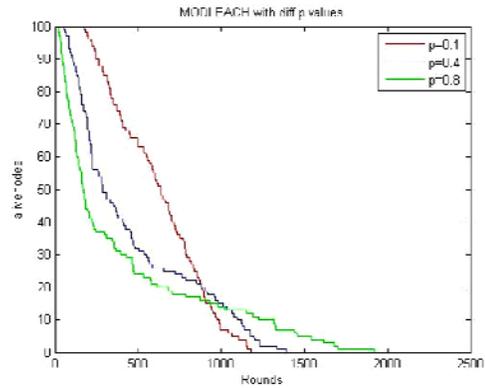

Fig: 1 Count of CHs with different p values

Fig: 2 Count of alive nodes with different p values

The graph represents the count of cluster heads with the variation in the value of $p$ (probability of choosing a CH). The plots clearly indicate that for the value of $p=0.1$, the CHs generated are too less whereas it is too high (approx reaching 90 in the first 200 rounds); a number too big which can contribute a lot to the consumption of energy; hence a nominal value for the selection of p for a reasonable generation of CHs is $p=0.4$.

The plots in fig: 2 reflect the number of alive nodes versus the number of rounds taken in a network for its completion or till the last node death. The plot is plotted varying the values of $p$. For $p=0.1$, the stability period (i.e. the death of the first node after the network starts its functionality) is quite better but the network dies quite early at around 1200 rounds. For $p=0.8$, the stability is quite less but the final rounds are going upto 1900 rounds, whereas for $p=0.4$, we get an intermediate, average and acceptable values in terms of stability period and maximum rounds traversed.

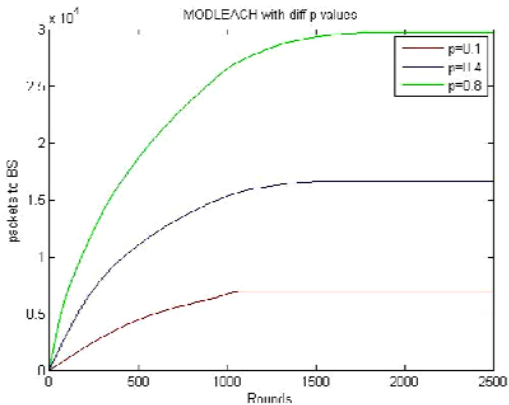
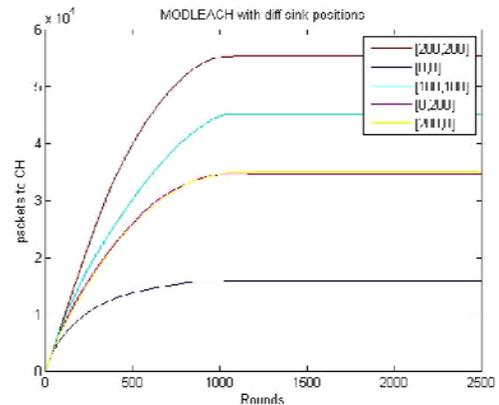

Fig: 3 Packets sent to BS with different p values

Fig: 4 Packets sent to CH with different sink locations

The plots in fig: 3 are plotted for different values of $p$ taking into consideration the packets sent to base station versus the rounds for which the network is working. The plots clearly reveal that with the growing value of $p$ from 0.1 till 0.8, the number of packets sent to base station are steadily growing and similar are the cases for the number of packets sent to cluster head from different nodes in a cluster with the rise in the values of $p$. But again there are trade-offs between different parameters involved with the variations in the value of $p$. Hence such value be chosen which can balance other parameters as well.

The plots in fig 4 are plotted for the packets transferred to cluster heads in various rounds of the network taking into consideration the various positions of the sink i.e. on the origin, on x-axis, on y-axis, in the middle of the network etc. Worst results are obtained when the sink is placed on the origin ($1.5 \times 10^4$ packets sent) and best results are obtained when placed in the middle of the network ($5.5 \times 10^4$). Rest plots lie in between these extreme ends.

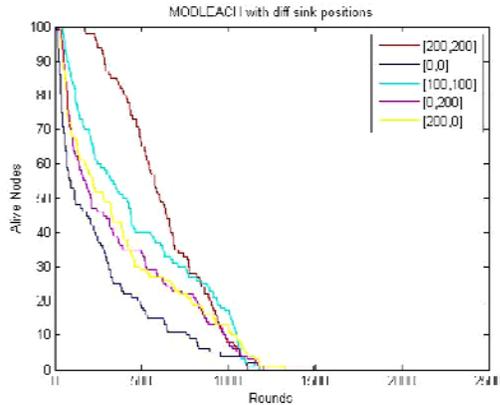 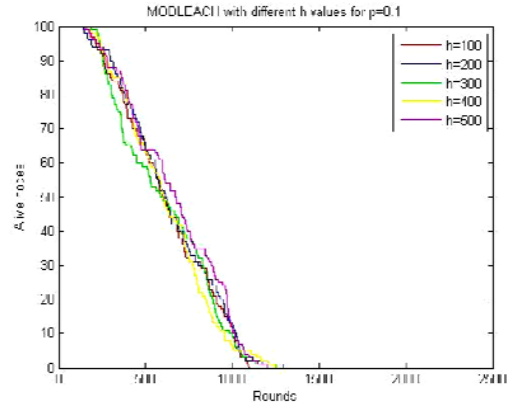

Fig: 5 Number of alive nodes at various sink locations     Fig: 6 Number of alive nodes for various values of *h*

Fig 5 indicates different plots for the alive nodes present in the network reflecting the stability of the network versus the maximum rounds traversed in consideration with the different sink positions. Again we find a trade-off between the stability of the network and the maximum rounds after which the network collapses. But this time, the variations are comparatively lesser as compared to those in fig 2 and the sink location in the middle of the network gives more promising results.

The plots in fig 6 are drawn by varying the value of hard threshold *h* keeping *p* as constant on some nominal value and noting the presence of alive nodes during the course of the network; but it is found that no major variations are found in the graphs of various quantities. Same experiment was also performed varying the values of soft threshold *s* and again the stability period was not found to vary much. Hence it is concluded that the soft and hard thresholds have no profound effect on the stability period of the network.

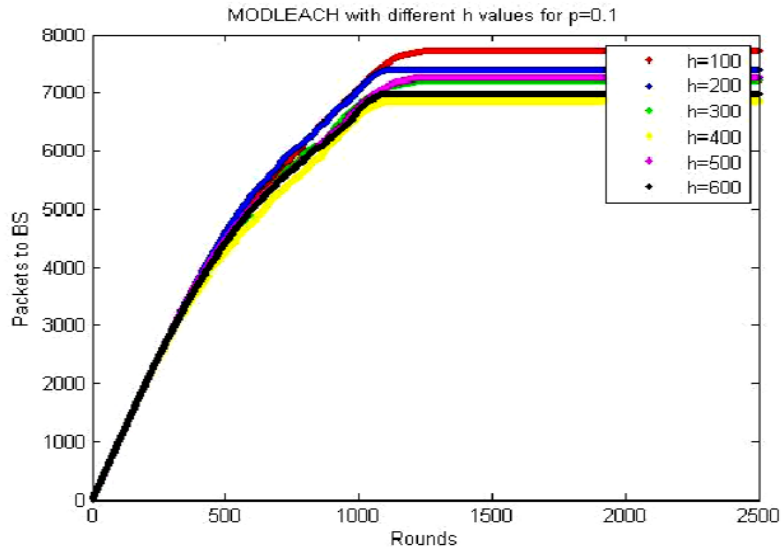

Fig: 7 Packets sent to base station with different *h* values

Again an experiment performed by keeping the value of *p* constant at 0.1 and varying the values of h and observing the variations in packets sent to the base station is shown graphically in fig 7. And much surprising results reveal that the values of packet sent to BS gradually decrease from *h*=100 to *h*=400, then again rise at *h*=500 and the repeated behaviour was observed from *h*= 500 to *h*=800. Hence a trade-off was again found between *p* and *h*. The experiment was repeated by changing the value of *p*=0.2 and then again varying the values of *h*.

**Conclusion and Future Work**

In this paper, we have proposed *i*MODLEACH protocol which is an extension to the MODLEACH protocol. Simulation results indicate that *i*MODLEACH outperforms MODLEACH in terms of network life-time and packets transferred to base station; that can further be utilized in other clustering routing protocols for better efficiency. Hence taking into consideration all these facts and figures, we can properly adjust the values of parameters *p*, *s* and *h* and certainly the performance of the *i*MODLEACH is considerably enhanced for different environments of applications.

In future, we are interested to deal the problem of energy efficiency at MAC layer like [17-23].